\documentclass[aps,pra,twocolumn,showpacs]{revtex4-1}

\usepackage{graphicx,amsmath,amssymb}

\begin{document}

\title{Quantum controlled-Z gate for weakly interacting qubits}
\author{Michal Mi\v{c}uda}
\affiliation{Department of Optics, Palack\' y University, 17. listopadu 1192/12,  771~46 Olomouc,  Czech Republic}

\author{Robert St\'{a}rek}
\affiliation{Department of Optics, Palack\' y University, 17. listopadu 1192/12,  771~46 Olomouc,  Czech Republic}

\author{Ivo Straka}
\affiliation{Department of Optics, Palack\' y University, 17. listopadu 1192/12,  771~46 Olomouc,  Czech Republic}

\author{Martina Mikov\'{a}}
\affiliation{Department of Optics, Palack\' y University, 17. listopadu 1192/12,  771~46 Olomouc,  Czech Republic}

\author{Miloslav Du\v{s}ek}
\affiliation{Department of Optics, Palack\' y University, 17. listopadu 1192/12,  771~46 Olomouc,  Czech Republic}

\author{Miroslav Je\v{z}ek}
\affiliation{Department of Optics, Palack\' y University, 17. listopadu 1192/12,  771~46 Olomouc,  Czech Republic}

\author{Radim Filip}
\affiliation{Department of Optics, Palack\' y University, 17. listopadu 1192/12,  771~46 Olomouc,  Czech Republic}

\author{Jarom\' ir Fiur\' a\v sek}
\affiliation{Department of Optics, Palack\' y University, 17. listopadu 1192/12,  771~46 Olomouc,  Czech Republic}

\begin{abstract}
We propose and experimentally demonstrate a scheme for the implementation of a maximally entangling quantum controlled-$Z$ gate between two weakly interacting systems. We conditionally enhance the
interqubit coupling by quantum interference. Both before and after the interqubit interaction, one of the qubits is coherently coupled to an auxiliary quantum system, 
and finally it is projected back onto qubit subspace. We experimentally verify the practical feasibility of this technique by using a linear optical setup with weak interferometric 
coupling between single-photon qubits. Our procedure is universally applicable to a wide range of physical platforms including hybrid systems such as atomic clouds or optomechanical oscillators coupled to light.
\end{abstract}


\maketitle

\section{Introduction}

Entangling two-qubit quantum gates are essential for universal quantum computing and quantum information processing \cite{Nielsen00}. 
However, the available interqubit coupling is often only weak and is limited by decoherence 
 \cite{Zurek03} or other factors. Moreover, in hybrid architectures connecting physically different qubits \cite{Hammerer10,optbook}, 
one can have only a limited amount of control over one of the systems. Schemes that would allow 
us to circumvent these obstacles and engineer highly entangling quantum gates under such unfavourable conditions are therefore highly desirable.

Here we propose and experimentally demonstrate a scheme for conditional implementation of a maximally entangling quantum controlled-$Z$ ($CZ$) gate between two qubits whose coupling 
can be arbitrarily weak. We show that the weak interqubit coupling can be enhanced by quantum interference. Both before and after the interqubit interaction, 
one of the qubits is coherently coupled to an auxiliary quantum level \cite{Lanyon09}, and finally it is projected back onto the qubit subspace. Remarkably, this procedure enhances the interqubit 
interaction strength although the coupling to auxiliary quantum level can be considered as a local bypass that allows the qubit to partly avoid interaction with the other qubit. 
Since this bypass is introduced only for one of the qubits, the scheme is suitable for hybrid architectures such as atomic clouds or optomechanical oscillators coupled 
to light \cite{Hammerer10,optbook}, where one of the systems is more difficult to address and manipulate.  We explicitly consider two important kinds of interactions: 
first, a conditional phase shift typical of spin-spin coupling, and, second, a beam-splitter type of coupling between two bosonic modes. 
We experimentally verify the practical feasibility of the proposed technique by using a linear optical setup with weak interferometric 
coupling between single-photon qubits \cite{Knill01,Kok07}.

The rest of the paper is organized as follows: In Sec. II we describe a protocol for conditional implementation 
of the quantum CZ gate between two weakly interacting atoms or spins. In Sec. III we show how to implemented the CZ gate between 
two weakly interferometrically coupled bosonic particles. Our experimental setup is described in Sec. IV and in Sec. V 
we present the experimental results. Finally, Sec. VI contains a brief summary and conclusions. Details of a theoretical model 
of our linear optical experimental setup are provided in the appendix.

\section{Weak spin-spin coupling} 

Consider first a spin-spin type of coupling between two spins, atoms or ions $A$ and $B$ \cite{CZneu,CZion,CZspin} that results in application of a controlled-phase gate
$U_\phi= \exp(i\phi|11\rangle\langle 11|)$ to qubits $A$ and $B$. 
As shown in Fig.~1, qubit A (B) is represented by two levels $|0\rangle$ and $|1\rangle$ of  particle $A$ ($B$), and our 
protocol also requires two additional levels $|2\rangle$ and $|3\rangle$ of particle $A$.  
In the computational basis $\{|00\rangle,|01\rangle,|10\rangle,|11\rangle\}$ the unitary matrix $U_\phi$ is diagonal and reads
\begin{equation}
U_\phi=\left(
\begin{array}{cccc}
1 & 0 & 0 & 0 \\
0 & 1 & 0 & 0 \\
0 & 0 & 1 & 0 \\
0 & 0 & 0 & e^{i\phi}
\end{array}
\right).
\end{equation}
Note that any two-qubit unitary operation diagonal in the computational basis is equivalent to $U_\phi$ up to unimportant local single-qubit phase shifts.
The phase shift $\phi$ provides a natural measure of the interaction strength and for $\phi=\pi$ we recover the quantum CZ gate, $U_{\mathrm{CZ}}=U_\pi$. 
An equivalent definition of the CZ gate is that it conditionally applies the sign flip operation $\sigma_Z=|0\rangle\langle 0|-|1\rangle\langle1|$ to one of the qubits
 if  the other qubit is in state $|1\rangle$.
 The CZ gate is equivalent to a CNOT gate up to single-qubit Hadamard transforms on the target qubit \cite{Nielsen00}.

A sequence of elementary operations leading to conditional implementation of the quantum CZ gate with arbitrary weak interqubit coupling $\phi$
is schematically illustrated in Fig.~1. To explain how the scheme works, we discuss how the sequence of elementary operations transforms a generic pure input two-qubit state 
\begin{equation}
|\Psi_{\mathrm{in}}\rangle_{AB}=c_{00}|00\rangle+c_{10}|10\rangle+c_{01}|01\rangle+c_{11}|11\rangle.
\end{equation}
As indicated in Fig.~1, we first drive the transition $|1\rangle_A \leftrightarrow |2\rangle_A$ to transfer part of population in state
$|1\rangle_A$ of qubit $A$ to another auxiliary level $|2\rangle_A$ of that particle, 
\begin{equation}
|1\rangle_A \rightarrow t|1\rangle_A +r|2\rangle_A, \quad |2\rangle_A \rightarrow t^*|2\rangle_A-r^*|1\rangle_A.
\end{equation}
Here $|t|^2+|r|^2=1$ and $t$ specifies the resulting strength of coupling of levels $|1\rangle_A$ and $|2\rangle_A$.
 After this preparatory step, the two particles interact as described by operation $U_\phi$, and their state becomes
\begin{eqnarray}
|\Psi_1\rangle_{AB}&=&c_{00}|00\rangle+tc_{10}|10\rangle+rc_{10}|20\rangle+c_{01}|01\rangle \nonumber \\
 & & +e^{i\phi} tc_{11}|11\rangle+rc_{11}|21\rangle.
\end{eqnarray}
After the interqubit coupling, the levels $|1\rangle_A$ and $|2\rangle_A$ of particle $A$ are coupled again, this time with a different coupling strength $\tilde{t}, \tilde{r}$. 
The state of particles $A$ and $B$ then reads
\begin{eqnarray}
|\Psi_2\rangle_{AB}&=&c_{00}|00\rangle+(t\tilde{t}-r\tilde{r}^\ast)c_{10}|10\rangle+(r\tilde{t}^\ast+t\tilde{r})c_{10}|20\rangle \nonumber \\
 & & +c_{01}|01\rangle +(e^{i\phi} t\tilde{t}-r\tilde{r}^\ast)c_{11}|11\rangle \nonumber \\ 
 & & +(r\tilde{t}^\ast+e^{i\phi}t\tilde{r})c_{11}|21\rangle.
\end{eqnarray}
Finally, we project particle $A$ onto the qubit subspace while also performing suitable coherent 
filtration to balance the states' amplitudes, $\Pi_{A}=\eta_A|0\rangle\langle 0|+|1\rangle\langle 1|$, where $\eta_A=t\tilde{t}-r\tilde{r}^*$.
Projection onto the subspace spanned by $\{|0\rangle_A,|1\rangle_A\}$ can be performed e.g., by driving a transition
between the auxiliary levels $|2\rangle_A$ and $|3\rangle_A$ of particle $A$ and conditioning on absence of fluorescence photons \cite{Leibfried03}.
After projection onto the qubit subspace, we can attenuate the amplitude of state $|0\rangle_A$  by coupling it to level $|2\rangle_A$ with coupling strength set to $t'=\eta_{A}$.
While state $|1\rangle_A$ is unaffected by such coupling, $|0\rangle_A$ is transformed to $t'|0\rangle_A+r'|2\rangle_A$. If we now project particle $A$ onto the qubit subspace,
we obtain the efective filtering transformation 
\begin{equation}
|0\rangle_A\rightarrow \eta_A|0\rangle_A, \qquad |1\rangle_A\rightarrow |1\rangle_A.
\end{equation} 
The final output state of particles $A$ and $B$ after successful projection onto qubit subspace and filtering is given by
\begin{eqnarray}
|\Psi_{\mathrm{out}}\rangle_{AB}&=&\eta_A c_{00}|00\rangle+\eta_A c_{10}|10\rangle +\eta_A c_{01}|01\rangle \nonumber \\
 & &  +(e^{i\phi} t\tilde{t}-r\tilde{r}^\ast)c_{11}|11\rangle.
\end{eqnarray}

\begin{figure}
\includegraphics[width=0.7\linewidth]{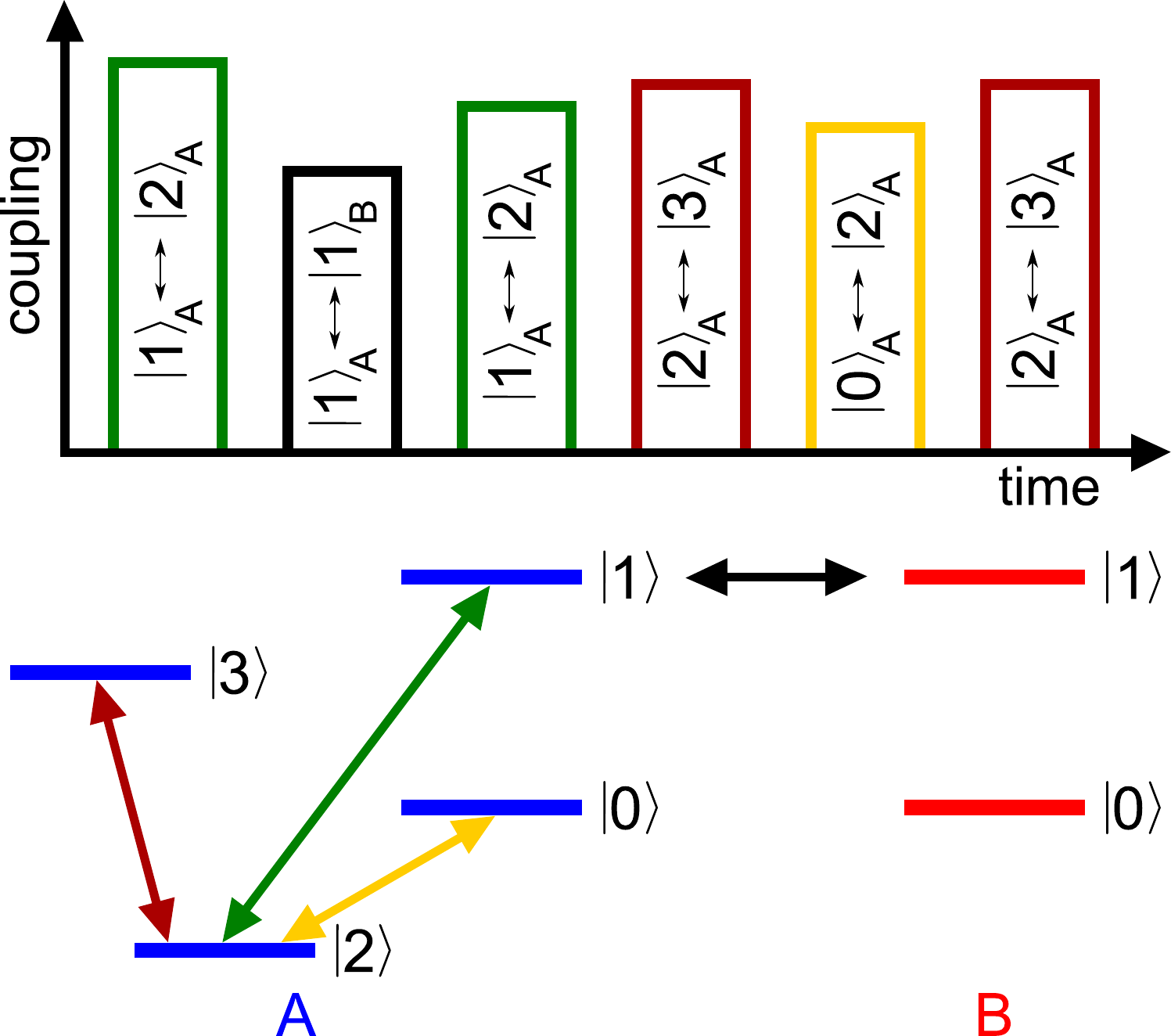}
\caption{(Color online)  Quantum CZ gate for weakly interacting single spins, ions, or atoms  whose coupling is described 
by a unitary operation $U_\phi$ with limited coupling strength $\phi$. Shown is the level scheme of the two particles $A$ and $B$, and a sequence 
of level couplings which allows to conditionally implement the CZ gate. For a detailed discussion, see text.}
\end{figure}

The resulting two-qubit operation $V$ defined as $|\Psi_{\mathrm{out}}\rangle= V |\Psi_{\mathrm{in}}\rangle$ is diagonal in the computational basis, 
\begin{equation}
V=\left(
\begin{array}{cccc}
\eta_A & 0 & 0 & 0 \\
0 & \eta_A & 0 & 0 \\
0 & 0 & \eta_A & 0 \\
0 & 0 & 0 & e^{i\phi}t\tilde{t}-r \tilde{r}^\ast
\end{array}
\right).
\end{equation}
The quantum CZ gate is conditionally implemented provided that $V=\eta_A U_{\mathrm{CZ}}$. This yields a single condition
$e^{i\phi}t\tilde{t}-r \tilde{r}^\ast=-\eta_A$, which is equivalent to 
\begin{equation}
\frac{r \tilde{r}^*}{t \tilde{t}}=\frac{1}{2}\left(1+e^{i\phi}\right). 
\label{trspin}
\end{equation}
For any $\phi$, this formula describes a whole parametric class of schemes and the coupling with level $|2\rangle$
can be optimized to maximize the success probability of the gate $P_S=|\eta_A|^2$. The optimal choice reads $|t|^2=|\tilde{t}|^2=1/[1+|\cos(\phi/2)|]$, which yields
\begin{equation}
P_S=\left(\frac{\sin(\phi/2)}{1+\left|\cos(\phi/2)\right|}\right)^{2}.
\end{equation}
Remarkably, the interplay of quantum interference and filtering \cite{Feizpour11,Simon11} enhances the interaction strength $\phi$ although the state $|2\rangle_A$
can be considered as a local bypass \cite{Lanyon09} that allows qubit $A$ partly avoid the interaction with qubit $B$.

\section{Weak interferometric coupling}
 
Let us now turn our attention to interferometric coupling of two bosonic modes described by a beam splitter
Hamiltonian $H_{\mathrm{BS}}=i\hbar \kappa (ab^\dagger-a^\dagger b)$,
where $a,b$ ($a^\dagger,b^\dagger$) denote annihilation (creation) operators of the two modes.
In the Heisenberg picture the annihilation operators are transformed according to
\begin{equation}
a_{\mathrm{out}}=t a_{\mathrm{in}}-rb_{\mathrm{in}},  \qquad b_{\mathrm{out}}=t b_{\mathrm{in}}+ra_{\mathrm{in}},
\label{BScoupling}
\end{equation}
 where $t=\cos(\kappa \tau)$  and $r=\sin(\kappa \tau)$ denote amplitude transmittance and reflectance, respectively, and $\tau$ is the effective interaction time. 
 We also introduce the intensity transmittance $T=t^2$ and reflectance $R=r^2$, which satisfy $T+R=1$.
Implementation of quantum gates based on beam splitter coupling has been intensively studied in the context of linear optics quantum computing \cite{Knill01,Kok07}, since
$H_{\mathrm{BS}}$ is the only practically available interaction between single photons. 
Although the resulting quantum gates are probabilistic by construction \cite{OBrien03,Gasparoni04,Politi08,Lanyon09},
their success probability can be increased by using auxiliary single photons and this architecture is in principle scalable \cite{Knill01}.

A linear optical CZ gate based on the beam splitter coupling of two photons \cite{Okamoto05,Langford05,Kiesel05,Lemr11} is depicted in Fig. 2(a).
 Each qubit is encoded into a state of a single photon that can propagate
in two modes labeled $A0$, $A1$, and $B0$, $B1$, respectively. A photon in mode $Q0$ ($Q1$)
represents a computational basis state $|0\rangle$ ($|1\rangle$) of qubit $Q=A,B$.
 The gate operates in the coincidence basis and a successful implementation of the gate is heralded by the presence 
 of a single photon in each pair of output modes $A0$, $A1$, and $B0$, $B1$.
The core of this linear optical CZ gate consists of a two-photon interference \cite{Hong87} on an unbalanced beam splitter BS with 
reflectance $R=r^2=2/3$, which occurs only if both qubits are in logical state $|1\rangle$. Conditional on the presence of a single
 photon in each pair of output modes, the coupling at the central beam splitter BS transforms the four basis states as follows
\begin{eqnarray}
|00\rangle &\rightarrow & |00\rangle, \nonumber \\ 
|01\rangle &\rightarrow & t|01\rangle, \nonumber \\ 
|10\rangle &\rightarrow & t|10\rangle, \nonumber \\ 
|11\rangle &\rightarrow & (t^2-r^2)|11\rangle.
\label{BSstates}
\end{eqnarray}
The auxiliary beam splitters BS$_A$ and BS$_B$ serve as quantum filters that attenuate the amplitudes of states $|0\rangle_A$ and $|0\rangle_B$. 
Assuming identical transmittances of all three beam splitters BS, BS$_A$, and BS$_B$, the conditional transformation of the four basis states reads
\begin{eqnarray}
|00\rangle &\rightarrow & T|00\rangle, \nonumber \\ 
|01\rangle &\rightarrow & T|01\rangle, \nonumber \\ 
|10\rangle &\rightarrow & T|10\rangle, \nonumber \\ 
|11\rangle &\rightarrow & (1-2R)|11\rangle.
\label{CZordinary}
\end{eqnarray}
The sign flip of the amplitude of state $|11\rangle$ required for the CZ gate occurs only if $R>1/2$ and the value $R=2/3$ 
is singled out by the condition $1-2R=-T=-(1-R)$ which ensures unitarity of the conditional gate (\ref{CZordinary}).
 The concept of linear optical quantum gates can be extended to other quantum systems exhibiting linear coupling between modes.
Of particular interest are implementations of hybrid gates between light and matter such as atomic ensembles \cite{Hammerer10} or mechanical oscillators \cite{optbook}.
In such a case, the achievable coupling strength $\kappa\tau$ is limited by decoherence so it may be impossible to directly achieve $R=2/3$
as required for the CZ gate.

\begin{figure}[!t!]
\includegraphics[width=\linewidth]{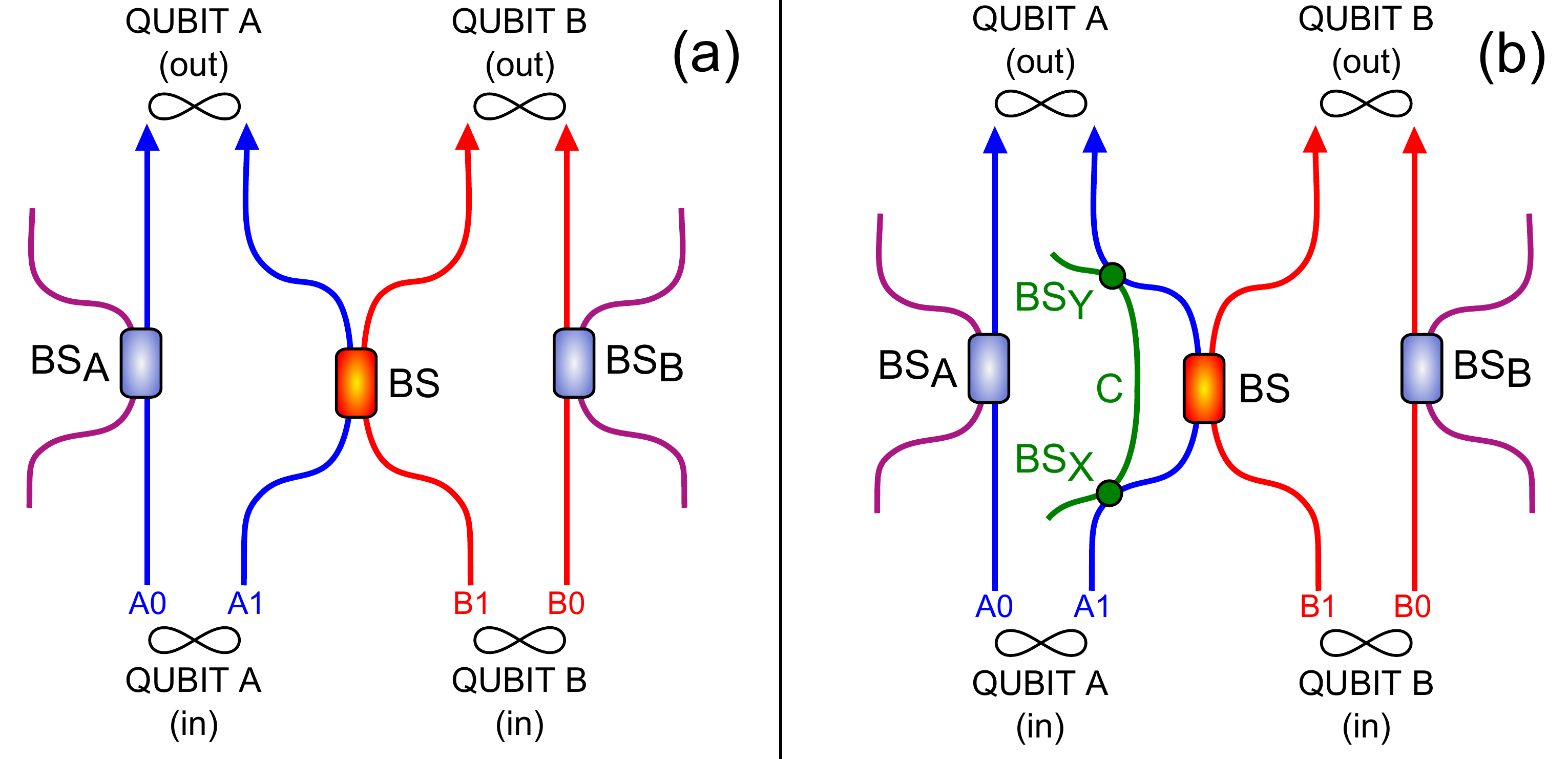}
\caption{
(a) (Color online) Linear optical quantum CZ gate 
operating in the coincidence basis. Qubits are encoded into paths of single photons. 
BS, BS$_A$ and BS$_B$ denote unbalanced beam splitters with intensity reflectance $R=2/3$. (b) CZ gate with arbitrary weak
interferometric coupling BS between two bosonic modes. Mode A1 is coupled to the bypass mode C by beam spitter couplings BS$_X$ and BS$_Y$.
Modes $A0$ and $B0$ are attenuated by coupling to auxiliary vacuum modes. In the hybrid realization, $B0$ and $B1$ schematically 
represent modes of an ensemble of atoms or a mechanical oscillator.}
\end{figure}

 In Fig. 2(b) we present a scheme that implements the quantum CZ gate with arbitrary weak linear coupling (\ref{BScoupling}) between two modes. 
 In analogy to the auxiliary level $|2\rangle_A$ considered in Section III, we introduce an auxiliary mode C to partly bypass the beam splitter interaction BS \cite{Lanyon09}.
 A {\em local} beam splitter coupling BS$_X$  with amplitude transmittance $t_X$ couples modes $A1$ and $C$.
After the beam splitter interaction between modes $A1$ and $B1$, mode $A1$ and bypass $C$ are {\em locally} recombined by a beam splitter interaction BS$_Y$ with amplitude transmittance $t_Y$.
 The scheme  also requires attenuation of modes $A0$ and $B0$, which can be accomplished by coupling them to auxiliary vacuum modes via beam-splitter interactions BS$_A$ and BS$_B$
 with amplitude transmittances $t_A$ and $t_B$, respectively. If we postselect on the presence of a single photon in each output port of the gate then the overall transformation $W$ 
 implemented by the setup in Fig. 2(b) is diagonal in the computational basis, $W|jk\rangle=w_{jk}|jk\rangle$, where
 \begin{eqnarray}
 w_{00} &=  & t_A t_B , \nonumber \\
 w_{01} & = &t_A t, \nonumber \\
w_{10} & = &\left(t t_X t_Y+r_Xr_Y\right)t_B, \nonumber \\
w_{11} & = &\left(2t^2-1\right)t_X t_Y+tr_Xr_Y.
\end{eqnarray}
 %where $r_j^2+t_j^2=1$.
The quantum CZ gate is conditionally implemented provided that $w_{00}=w_{10}=w_{01}=-w_{11}$, which yields the conditions,
 $t_A=tt_Xt_Y+r_Xr_Y$, $t_B=t$,  and 
 \begin{equation}
 \frac{r_Xr_Y}{t_Xt_Y}=\frac{3R-2}{2t},
 \end{equation}
 which is analogous to condition (\ref{trspin}) derived for the spin-spin coupling. 
 The success probability of the gate $P_S=R^2 t_X^2t_Y^2/4$ is maximized when $t_X^2=t_Y^2=2t/(2t+|3R-2|)$.
The quantum interference conditionally enhances the coupling of modes $A1$ and $B1$ although this interaction is
 partially bypassed by coupling mode A1 with mode C. Since the bypass is introduced only for one system,
 the scheme is suitable for hybrid architectures \cite{Hammerer10,optbook} where one of the systems is more difficult to address and manipulate.

\section{Experimental setup}

We have experimentally demonstrated the weak-coupling enhancement with a linear optical setup where photons interfere
on a beam splitter with $R=1/3$, see Fig.~3. This platform provides a suitable testbed for verifying the practical feasibility of our method 
and checking its robustness against various experimental imperfections.
We utilize time-correlated photon pairs generated by the process of collinear frequency-degenerate type-II spontaneous parametric
down-conversion in a 2 mm thick BBO crystal pumped by a continuous-wave laser diode at 405 nm \cite{Micuda14}.
 Basis states $|0\rangle$ and $|1\rangle$ are represented by horizontally and vertically polarized input photons, respectively.
An arbitrary polarization state of each photon can be prepared using a sequence of quarter- and half-wave plates.
The polarization state of the signal photon is converted to path encoding with the use of a calcite beam displacer
that introduces a transversal spatial offset between horizontal and vertical polarizations.
Qubit $A$ is thus encoded into the path of a signal photon in a Mach-Zehnder interferometer formed by two calcite beam displacers \cite{Lanyon09,Micuda13}, 
and qubit $B$ is represented by polarization of the idler photon. Since $1-2R=1/3>0$, we are deep in the weak-coupling regime.

\begin{figure}[!t!]
\centerline{\includegraphics[width=\linewidth]{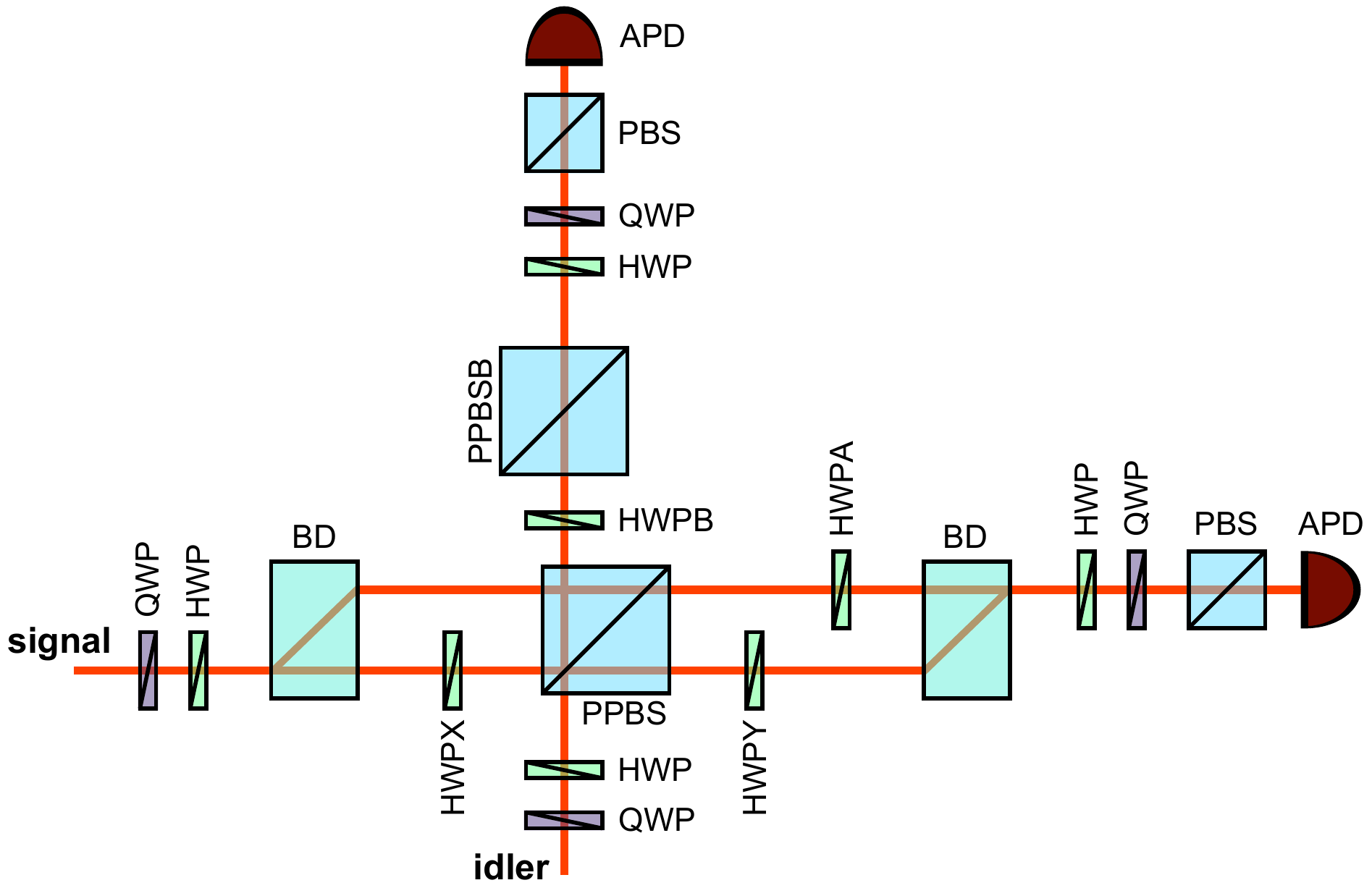}}
\caption{(Color online) Experimental setup. HWP - half-wave plate, QWP - quarter-wave plate, 
PPBS - partially polarizing beam splitter with reflectances $R_V=1/3$ and $R_H=0$ for vertical and horizontal polarizations, respectively,
PBS - polarizing beam splitter, BD - calcite beam displacer, APD - single-photon detector.}

\end{figure}

\begin{figure}[!b!]
\centerline{\includegraphics[width=\linewidth]{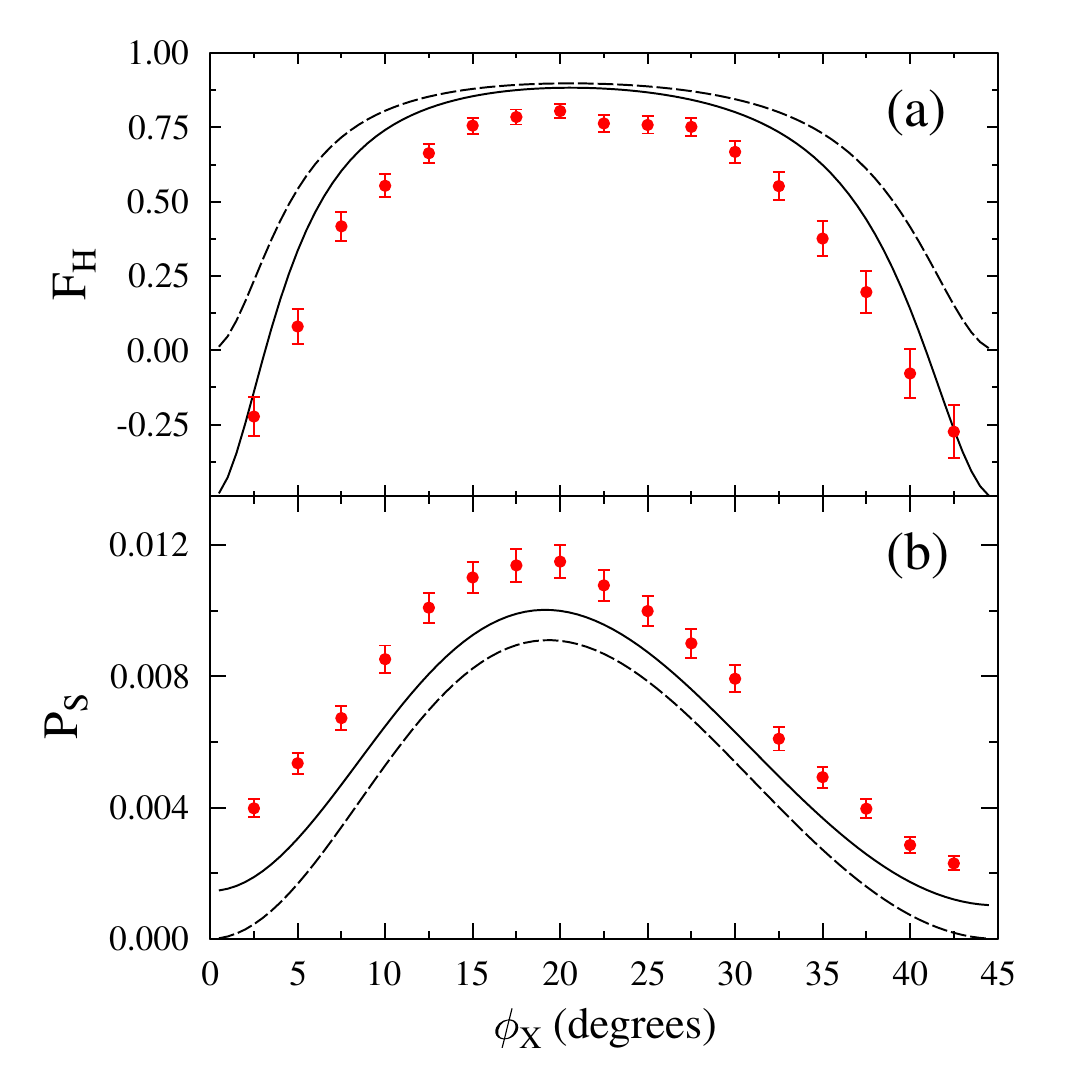}}
\caption{(Color online) (a) Hofmann lower bound $F_H$ on gate fidelity and (b) success probability of the gate $P_S$
are plotted as functions of coupling to the bypass $\phi_X$. Circles represent experimental data and solid lines indicate predictions of theoretical model.
Error bars represent statistical errors ($3\sigma$) that were determined assuming Poissonian statistics of coincidence counts.
 The dashed line in panel (a) shows the actual gate fidelity $F_\chi$ as predicted by the theoretical model. 
The dashed line in panel (b) represents success probability of an ideal scheme with $\mathcal{V}=1$ and $R_H=0$.}
\end{figure}

\begin{figure*}[!t!]
\centerline{\includegraphics[width=\linewidth]{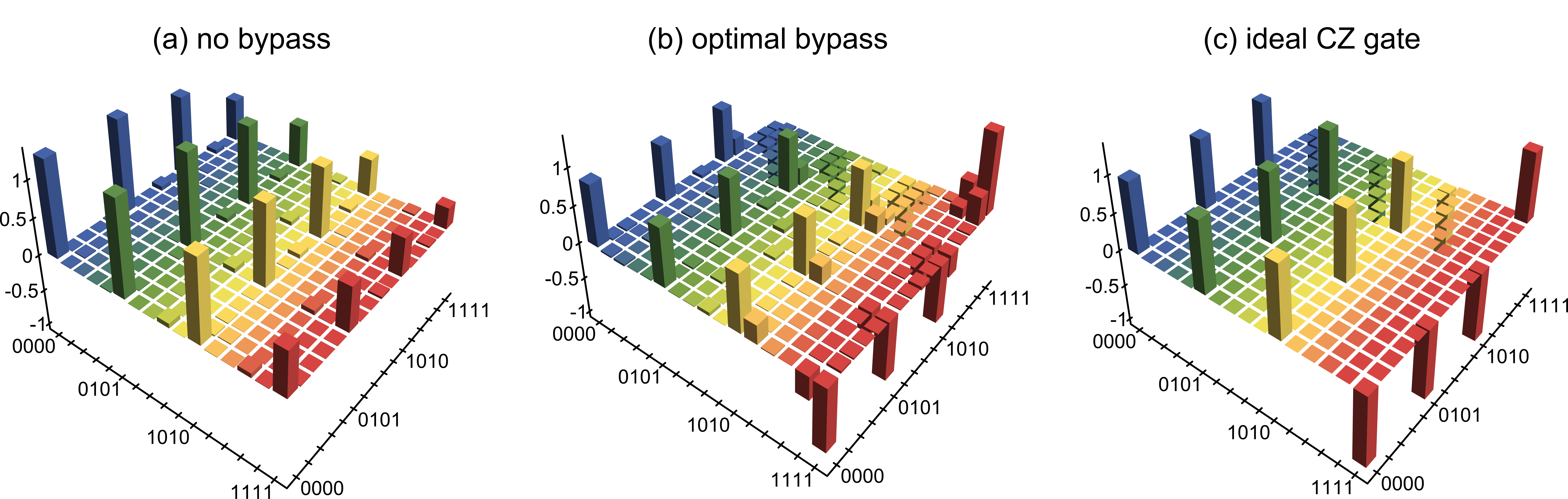}}
\caption{(Color online) Quantum process tomography of the CZ gate. The figure displays the reconstructed quantum process matrix $\chi$ of the two-qubit operation without bypass
(a) and with optimal bypass (b). For comparison, the ideal CZ gate is shown in panel (c). For ease of comparison, all process matrices are normalized so that
$\mathrm{Tr}[\chi]=4$. The experimentally determined success probability of the CZ gate shown in panel (b) reads $P_{S}=1.15(2)\%$.}
\end{figure*}

 The bypass is implemented with the use of the polarization degree of freedom of the signal photon. 
Specifically, modes A1 and C correspond to the vertically and horizontally polarized modes propagating in the lower arm of the interferometer, 
and their coupling is provided by two half-wave plates HWPX and HWPY. Attenuation of mode A0 is achieved by rotation of
the half-wave plate HWPA, because the second beam displacer BD2 transmits only a horizontally polarized signal from the upper interferometer arm to the output. Mode B0 
 is attenuated by a combination of a partially polarizing beam splitter PPBSB and a half-wave plate HWPB rotated by $45^\circ$, which swaps the horizontal and vertical polarizations.
 At the gate output each qubit can be measured in an arbitrary basis by using a combination
 of quarter- and half-wave plate, polarizing beam splitter, and a single-photon detector APD. The number of coincidence clicks of the two detectors in a fixed detection time of $10$~s
 was measured for various input states and measurement bases.

\section{Experimental results}

We have experimentally characterized the performance of the CZ gate by a Hofmann lower bound \cite{Hofmann05,Micuda14} on the quantum gate fidelity $F_\chi$   \cite{Horodecki99}, which
 is defined as an overlap of the Choi matrices \cite{Choi75,Jamiolkowski72} $\chi$ and $\chi_{\mathrm{CZ}}$ of the actual two-qubit operation 
and the ideal CZ gate, respectively, 
\begin{equation}
F_{\chi}=\frac{\mathrm{Tr}[\chi\chi_{\mathrm{CZ}}]}{\mathrm{Tr}[\chi]\mathrm{Tr}[\chi_{\mathrm{CZ}}]}.
\end{equation}
For two-qubit operations, $\chi$ is a positive semidefinite operator on a four-qubit Hilbert space. In particular,
$\chi_{\mathrm{CZ}}=|\chi_{\mathrm{CZ}}\rangle\langle \chi_{\mathrm{CZ}}|$ is a density matrix of a pure maximally entangled state,
\begin{equation}
|\chi_{\mathrm{CZ}}\rangle=|0000\rangle+|0101\rangle+|1010\rangle-|1111\rangle.
\end{equation}
The Hofmann lower bound $F_H$ on the gate fidelity $F_\chi$ is determined by average state fidelities $F_1$ and $F_2$ for two mutually unbiased bases \cite{Hofmann05},
\begin{equation}
F_\chi\geq F_1+F_2-1=F_H.
\end{equation}
Let $\{|\psi_{jk}\rangle\}_{k=1}^4$ denote the states forming the $j$th probe basis. The average state fidelity $F_{j}$ is defined as \cite{Micuda14}
\begin{equation}
F_j=\frac{\sum_{k=1}^4 p_{jk} f_{jk}}{\sum_{k=1}^4 p_{jk}},
\end{equation}
where $p_{jk}$ denotes the success probability of the gate for input state $|\psi_{jk}\rangle$, and $f_{jk}$ denotes the fidelity 
of the corresponding normalized output state with the ideal output $U_{\mathrm{CZ}}|\psi_{jk}\rangle$.
In our experiment, we employed the following two mutually unbiased product bases  $\{|0+\rangle,|0-\rangle,|1+\rangle,|1-\rangle\}$ and
$\{|\!+\!0\rangle,|\!+\!1\rangle,|\!-\!0\rangle,|\!-\!1\rangle\}$, where $|\pm\rangle=\frac{1}{\sqrt{2}}(|0\rangle\pm |1\rangle)$. 
Since the quantum CZ gate maps all these input states onto product states, the output-state fidelities can be directly measured \cite{Okamoto05,Micuda14}.
 Besides a bound on gate fidelity, we characterize the gate operation  also by its average success probability, which is defined as 
\begin{equation}
P_S=\frac{1}{4} \sum_{k=1}^4p_{jk}.
\label{PSdef}
\end{equation}
As shown in the appendix, $P_S$ does not depend on the choice of the basis.   

The measurement of the Hofmann bound served for a fast and efficient characterization of our scheme for a wide range of values of the rotation angle $\phi_X$ of HWPX
 which specifies the strength of coupling to the auxiliary mode $C$, $t_X=\cos(2\phi_X)$. The practical advantage of the Hofmann bound is that its determination 
 requires much less experimental settings than an exact estimation of the gate fidelity. This is important in our case because each change of measurement setting 
 lasts about $6$~s on average, 
 determined by the speed of the motorized wave-plate rotations. With measurement time of each coincidence set to $10$~s, the whole measurement lasted less than $3$ hours.
  We were thus able to collect all data  
 in a single measurement run during which the interferometer remained passively stable. 
 The data were thus recorded under identical conditions and are mutually comparable.
 By contrast, we estimate that in case of determination of the true gate fidelity the time required for changes of the wave plate settings would be 
 about 20 minutes for each value $\phi_X$, hence more than $5$ hours in total for the $17$ values of $\phi_X$ that we probed.

The experimentally determined dependence of the Hofmann fidelity bound $F_H$ and the success probability of the gate $P_S$ on $\phi_X$ is  plotted in Fig. 4.
 The experimental data are shown as red circles, while the solid lines represent theoretical predictions of $F_H$ and $P_S$ based  
on a model which is described in detail in the Appendix. This model takes into account imperfect two-photon interference
with visibility $\mathcal{V}=0.94$ and the imperfection of the central partially polarizing beam splitter PPBS which exhibited a small nonzero reflectance
for horizontally polarized photons, $R_H=0.019$, while ideally it should be perfectly transmitting for horizontally polarized light. 
The model correctly predicts the behaviour of both $F_H$ and $P_S$.
Slight differences between the theory and the experimental data can be attributed to imperfections of wave plates and other optical componets and residual 
long-term phase fluctuations in the calcite interferometer, which are not included in our model. The optimal operating point $\phi_X=20^\circ$ 
where the gate achieves maximum fidelity coincides with the point where the gate also achieves maximum success probability.
At this point, the parasitic coupling due to nonzero $R_H$ is least harmful. As the success probability decreases this coupling becomes more
significant and eventually it completely alters the gate behavior. Note that under ideal conditions, the gate operation should
be perfect and $F_H=F_\chi=1$ should hold for any $\phi_X$ and only the success probability would depend on $\phi_X$.

 At the optimal operating point we have carried out a complete quantum process tomography \cite{Paris04,OBrien04b} of the CZ gate.
 Each input qubit was prepared in six states  $\{|0\rangle,|1\rangle,|+\rangle,|-\rangle,|r\rangle,|l\rangle\}$,
and each output qubit was measured in three bases $\{|0\rangle,|1\rangle\}$, $\{|+\rangle,|-\rangle\}$, $\{|r\rangle,|l\rangle\}$,
 where $|r\rangle=\frac{1}{\sqrt{2}}(|0\rangle+i |1\rangle)$, and $|l\rangle=\frac{1}{\sqrt{2}}(|0\rangle -i |1\rangle)$. 
 The Choi matrix $\chi$ was then reconstructed from this data by using a Maximum Likelihood estimation algorithm \cite{Micuda14}.
The results are plotted in Fig. 5. For comparison, we have first set $t_X=t_Y=t_A=1$, i.e. the bypass was not used. The resulting operation
is shown in Fig. 5(a) and it resembles the identity operation up to some amplitude attenuation. In contrast, when
we operate the bypass at the optimal working point $\phi_X= 20^\circ$,
we obtain the operation shown in Fig. 5(b) which closely resembles the ideal CZ gate shown in Fig. 5(c).
The change of sign of the state $|11\rangle$ is clearly visible in the figure as negative off-diagonal elements of $\chi$,
and the fidelity of this operation with CZ gate reads $F=0.846$. This is consistent with our theoretical
model that predicts $F_{\chi}=0.889$, see also the dashed line in Fig. 4(a).

\section{Conclusions}

In summary, we have proposed and experimentally verified a procedure for conditional enhancement of a weak interqubit interaction.
We have shown that the weak interaction can be enhanced by a combination of coupling to auxiliary modes or internal quantum levels \cite{Lanyon09}, 
quantum interference, and filtering and we have demonstrated that this procedure allows us to implement a maximally entangling 
quantum controlled-$Z$ gate between weakly coupled qubits. Our proof-of-principle experimental implementation with linear 
optics verified that the method is experimentally feasible and sufficiently robust. We envision that this technique may be particularly suitable for implementation of highly entangling
quantum gates in hybrid quantum information processing architectures, where light is coupled to matter and the achievable interaction strength
 is limited by decoherence or other effects. We hope that our work will stimulate further investigations of the applicability of our technique to such architectures.

\begin{acknowledgments}

This work was supported by the Czech Science Foundation (13-20319S). R.S. acknowledges support by Palacky University (IGA-PrF-2014008 and IGA-PrF-2015-005).

\end{acknowledgments}

\appendix*
\section{Theoretical model of the linear optical CZ gate}

Here we describe a  theoretical model of the linear optical quantum CZ gate with weak coupling shown in Fig.~3 of the main text. 
The model takes into account imperfect visibility $\mathcal{V}$ of two-photon interference on the central partially polarizing beam splitter PPBS, 
and also the fact that this beam splitter is partially reflecting also for horizontally polarized photons \cite{Nagata09,Micuda14}.
We denote the amplitude transmittance and reflectance of PPBS for horizontally polarized photons by $t_H$ and $r_H$, respectively. 
Since both partially polarizing beam splitters PPBS and PPBSB were manufactured in a single batch, 
we shall assume that their parameters are identical, which is consistent with independent measurements of their transmittances and reflectancies, that yielded 
$R=r^2=0.313$, and $R_H=r_H^2=0.019$. Note that, due to various manufacturing imperfections, the actual reflectance of PPBS for vertically polarized photons 
 differed slightly from the nominal value of $1/3$. The amplitude transmittances $t_A$, $t_X$, $t_Y$ and reflectances $r_X$, $r_Y$ are related 
to the rotation angles $\phi_A$, $\phi_X$ and $\phi_Y$  of half-wave plates HWPX, HWPY, and HWPA as $t_j=\cos(2\phi_j)$ and $r_j=\sin(2\phi_j)$, $j=A,X,Y$.

Imperfect visibility of two-photon interference $\mathcal{V}=0.94$ is accounted for by assuming that with probability
\begin{equation}
q=\frac{2\mathcal{V}}{1+\mathcal{V}}
\end{equation}
the two photons are indistinguishable and perfectly interfere, while with probability $1-q$ the two photons are perfectly distinguishable \cite{Micuda14}. 
Conditional on presence of a single photon in each output port of the gate, the resulting two-qubit operation $\chi$ then becomes a mixture of three contributions,
\begin{equation}
\chi=q\chi_{I}+(1-q)\chi_{R}+(1-q)\chi_{T},
\end{equation}
where 
\begin{equation}
\chi_{I}=|\chi_{I}\rangle \langle \chi_{I}|, \qquad
\chi_{R}=|\chi_{R}\rangle \langle \chi_{R}|, \qquad
\chi_{T}=|\chi_{T}\rangle \langle \chi_{T}|.
\end{equation}
The first term $\chi_{I}$ corresponds to interference of indistinguishable photons, while $\chi_R$ and $\chi_T$ represent 
the transformations when the photons are distinguishable and are both either
reflected or transmitted at the central PPBS. Below we specify the resulting transformations of input computational basis states for each of these contributions and we 
provide explicit expressions for the process matrices $\chi_{I}$, $\chi_{R}$, and $\chi_{T}$.

Let us first consider the case of indistinguishable photons. The resulting transformation of the computational basis states reads
\begin{equation}
\begin{array}{lcl}
|00\rangle \rightarrow \beta_{00}|00\rangle, & \qquad &  |10\rangle \rightarrow \beta_{10}|10\rangle+\gamma_{11}|11\rangle, \\[1mm]
|01\rangle \rightarrow \beta_{01}|01\rangle, & \qquad & |11\rangle \rightarrow \beta_{11}|11\rangle+\gamma_{10}|10\rangle,
\end{array}
\label{transformationI}
\end{equation}
where
\begin{equation}
\begin{array}{l}
\beta_{00}=\beta_{01}=t t_A t_H^2,\\[1mm]
\beta_{10}=t_X t_Y t_H t^2+ r_X r_Y(t_H^2-r_H^2)t, \\[1mm]
\beta_{11}=t_Xt_Y t_H(2t^2-1)+r_Xr_Y t_H^2 t,\\[1mm]
\gamma_{11}=-r_H t_X r_Y r t_H, \\[1mm]
\gamma_{10}=- t r r_X t_Y r_H.
\end{array}
\end{equation}
Note that due to the presence of parasitic coupling $r_H$, the computational basis states $|10\rangle$ and $|11\rangle$ are mapped onto superpositions of these states, so the gate no longer preserves the form
$|jk\rangle \rightarrow \beta_{jk}|jk\rangle$. The process matrix $\chi_I$ corresponding to transformation (\ref{transformationI}) can be expressed as follows,
\begin{eqnarray}
|\chi_I\rangle &= &\beta_{00}|00\rangle|00\rangle+\beta_{01}|01\rangle|01\rangle+\beta_{10}|10\rangle|10\rangle \nonumber \\
& & +\beta_{11}|11\rangle|11\rangle+\gamma_{11}|10\rangle|11\rangle+\gamma_{10}|11\rangle|10\rangle. \nonumber \\
\end{eqnarray}
Let us next assume that the photons are distinguishable and  both transmitted through the central PPBS. In this case, the resulting transformation reads
\begin{eqnarray}
|00\rangle &\rightarrow& \beta_{00}^T|00\rangle, \qquad |10\rangle \rightarrow \beta_{10}^T|10\rangle, \nonumber  \\
|01\rangle &\rightarrow& \beta_{01}^T|01\rangle, \qquad |11\rangle \rightarrow \beta_{11}^T|11\rangle,
\label{trT}
\end{eqnarray}
where
\begin{eqnarray}
\beta_{00}^T=\beta_{01}^T&=&  t_H^2 t t_A , \nonumber \\
\beta_{10}^T=\beta_{11}^T&=& t_H t(t_X t_Y t + r_X r_Y t_H).
\end{eqnarray}
The process matrix $\chi_T$ corresponding to transformation (\ref{trT}) reads,
\begin{equation}
|\chi_T\rangle= \beta_{00}^T|00\rangle|00\rangle+\beta_{01}^T|01\rangle|01\rangle+\beta_{10}^T|10\rangle|10\rangle+\beta_{11}^T|11\rangle|11\rangle.
\end{equation}
Finally, we consider distinguishable photons that are both reflected at the central PPBS. In this case, the resulting transformation reads
\begin{eqnarray}
|10\rangle &\rightarrow& \beta_{10}^R|10\rangle+\gamma_{11}^R|11\rangle, \nonumber \\
|11\rangle &\rightarrow& \beta_{11}^R|11\rangle+ \gamma_{10}^R|10\rangle,
\end{eqnarray}
where
\begin{equation}
\begin{array}{lcl}
\beta_{10}^R= -r_H^2 r_X r_Y t, &\qquad & \gamma_{11}^R=-r_H t_X r_Y r t_H, \\[1mm]
\beta_{11}^R= -r^2 t_X t_Y t_H,  &\qquad & \gamma_{10}^R=-r r_X t_Y r_H t.
\end{array}
\end{equation}
No output is produced for input states $|00\rangle$ and $|01\rangle$ in this case, hence we have
\begin{equation}
|\chi_R\rangle= \beta_{10}^R|10\rangle|10\rangle+\beta_{11}^R|11\rangle|11\rangle+\gamma_{11}^R|10\rangle|11\rangle+\gamma_{10}^R|11\rangle|10\rangle.
\end{equation}

With the explicit expression for the quantum process matrix of the gate at hand, we can calculate the state fidelities $f_{jk}$ and success probabilities $p_{jk}$ required for determination of the Hofmann bound $F_H$.
The operation $\chi$ transforms an input density matrix $\rho_{\mathrm{in}}$ as follows \cite{Fiurasek01},
\begin{equation}
\rho_{\mathrm{out}}= \mathrm{Tr}_{\mathrm{in}}[\rho_{\mathrm{in}}^T \otimes I \, \chi],
\end{equation}
where $T$ denotes transposition in the computational basis, $\mathrm{Tr}_{\mathrm{in}}$ represents partial trace over the input Hilbert space, 
 and $I$ denotes the identity operator on the output Hilbert space. 
The success probability of operation $\chi$ for input $\rho_{\mathrm{in}}$ is given by a trace of the output density matrix.
For a pure input state $|\psi_{jk}\rangle$ we have 
\begin{equation}
p_{jk}=\mathrm{Tr}[\psi_{jk}^T \otimes I \chi],
\end{equation}
where $\psi_{jk}=|\psi_{jk}\rangle\langle \psi_{jk}|$ is a short-hand notation for a density matrix of a pure state. 
The state fidelity $f_{jk}$ is defined as a normalized overlap between the actual output state $\mathrm{Tr}_{\mathrm{in}}[\psi_{jk}^T\otimes I \, \chi]$ and the ideal output of the CZ gate,
$U_{\mathrm{CZ}} |\psi_{jk} \rangle $. This yields 
\begin{equation}
f_{jk}= \frac{\mathrm{Tr}[\psi_{jk}^T\otimes U_{\mathrm{CZ}}\psi_{jk} U_{\mathrm{CZ}}^\dagger \,\chi]}{\mathrm{Tr}[\psi_{jk}^T\otimes I \, \chi]}.
\end{equation}
The average success probability of the gate $P_S$ defined in Eq. (\ref{PSdef}) can be expressed as
\begin{equation}
P_S=\frac{1}{4}\sum_{k=1}^4 p_{jk}=\frac{1}{4}\sum_{k=1}^4\mathrm{Tr}[\psi_{jk}^T \otimes I \,\chi ]=\frac{1}{4}\mathrm{Tr}[\chi],
\end{equation}
where we used the identity $\sum_{k=1}^4 \psi_{jk}=I$, which is valid for any basis.
This proves that $P_S$ does not depend on the choice of the basis, as claimed in the main text.

\end{document}